# THE VATICAN AND THE FALLIBILITY OF SCIENCE:
# AUGUSTINE, COPERNICUS, DARWIN AND RACE[1]

Christopher M. Graney
Vatican Observatory
c.graney@vaticanobservatory.org

ABSTRACT

**This paper provides an overview of work, published since the opening of the archives of the Vatican Congregation for the Doctrine of the Faith at the end of the twentieth century, regarding the Vatican confronting evolution in the nineteenth century. It argues that this work, considered in light of recent studies of scientific writings by Jesuit astronomers who in the seventeenth century were opposed to the ideas of Copernicus, points to interesting things yet to be learned regarding the Vatican's actions on heliocentrism. Concern for Scripture and for the fallible and consequential nature of science, together with the processes used by the Vatican in these confrontations, inevitably led to messy results in these well-known "religion and science" confrontations. Nevertheless, these confrontations suggest that what the Vatican was attempting to do in confronting evolution or heliocentrism is something that is needed in science, and something that will be done in the future, probably not by the Vatican, and probably in a fashion not less messy.**

INTRODUCTION

How did the Vatican handle science in the early seventeenth century, when its Congregation of the Index rejected heliocentrism, Nicolas Copernicus's idea that the Earth circles the Sun, as "false" and "altogether contrary to Holy Scripture"?[2] That rejection is a central event in the histories of religion and of science. Some light has been shed on this question inadvertently by studies of how the Vatican handled science in the late nineteenth century, when the Congregation of the Index dealt with the idea of evolution, and especially by studies made after the opening of the archives of the Congregation for the Doctrine of the Faith (formerly the Holy

---

[1] For conference on "Unity & Disunity in Science: Philosophical, Historical, and Theological Perspectives (Spring 2024)", University of Notre Dame, April 4-6, 2024: https://sites.nd.edu/hpstv-conference/home/unity-disunity-in-science-philosophical-historical-and-theological-perspectives-spring-2024/

[2] "Decree of the Index (March 5, 1616)," in *The Galileo Affair: A Documentary History*, Maurice A. Finocchiaro ed. & trans. (Berkeley, California: University of California Press, 1989), 149.

Office) at the end of the twentieth century. The studies considered here are those by Barry Brundell; by Mariano Artigas, Thomas E. Glick, and Rafael A. Martinez; and by John P. Slattery.[3]

The connection between the questions of heliocentrism and of evolution seems natural. Artigas, Glick and Martinez (henceforth sometimes abbreviated AG&M) open the first chapter of their book *Negotiating Darwin: The Vatican Confronts Evolution, 1877-1902* by stating that "the cases of Galileo and of evolution have become emblematic of the problems between science and religion"; they close their last chapter with mention of how "the shadow of Galileo surely fell over the evolution controversy".[4] Galileo features prominently in Slattery's discussion.[5] Brundell writes of Church authorities being "anxious to avoid another Galileo case" when dealing with evolution.[6]

That natural connection may help us understand the actions of those authorities in the heliocentrism case, through a comparison with what happened in the evolution case where so much information is available. New studies regarding that "shadow of Galileo", and in particular regarding the role of Jesuit astronomers in the heliocentrism controversy, suggest that how the Vatican confronted evolution provides clues to how it confronted heliocentrism. Brundell, AG&M and Slattery all reveal a Vatican confronting the science of evolution through processes that seem haphazard and inefficient on one hand, yet engaged with the still-unsettled science on the other. Those processes likely echoed what happened when the Vatican confronted heliocentrism.

Such confrontations appear messy, especially when seen from the vantage point of today. They may be inevitable, however. They may even be needed, given those haphazard human processes and the consequential yet fallible nature of science.

VATICAN PROCESSES: CONGREGATIONS, CONSULTORS, AND REPORTS

Artigas, Glick and Martinez provide a detailed study of the Vatican confronting the science of evolution and of the processes involved in that confrontation. They treat six cases that arose in the late nineteenth century from Catholics writing on Darwin's theory of evolution. Slattery's book focuses on one of those six cases, namely that of Fr. John Zahm, CSC. Three of its five chapters are devoted to Zahm's biography and training and to general trends in Catholic thought over the nineteenth century, leaving the lesser portion of the book to the matter of the Vatican and Zahm. Brundell's journal article is devoted mostly to the Zahm case, but touches on other cases as well.

---

[3] Barry Brundell, "Catholic Church Politics and Evolution Theory, 1894-1902", *The British Journal for the History of Science* 34:1 (2001), 81-95; Mariano Artigas, Thomas E. Glick, and Rafael A. Martinez, *Negotiating Darwin: The Vatican Confronts Evolution, 1877-1902* (Baltimore, Maryland: Johns Hopkins University Press, 2006); Brundell, "Book Reviews: *Negotiating Darwin*", *Isis* 99:2 (2008), 415-16. John P. Slattery, *Faith and Science at Notre Dame: John Zahm, Evolution, and the Catholic* Church (Notre Dame, Indiana: University of Notre Dame Press, 2019). These are authors Slattery discusses as having published following the 1998 opening of the Archives for the Congregation for the Doctrine of the Faith for scholarly research (18-19).
[4] *Negotiating Darwin*, 7, 281.
[5] *Faith and Science at Notre Dame*, 78-82.
[6] "Books Reviews", 416.

Artigas, Glick and Martinez point out that as the Vatican confronted evolution it issued no official pronouncements or judgements on the subject. Nevertheless, they say, many individuals, both within the Vatican and without, had opinions on the subject, and wrote and took actions accordingly. AG&M note how—because there were so many opinions and so much written and done, and yet so little official action—over the past century various secondary and tertiary writers have stepped forward to discuss, without first-hand information, what the Vatican was "really" saying about Darwin and evolution. AG&M emphasize the archival materials that the Vatican opened to scholars in 1998, and that these materials show that those secondary and tertiary writers have often been mistaken or imprecise in assessments which were then passed on, especially through textbooks, to "generations of seminarians, priests, and professors the world over" who "studied Catholic theology textbooks in which it was explained that the Holy See had acted against evolutionism in these cases". Thus the authors emphasize "the need to present the new archival data objectively. Only in this way is it possible to get beyond erroneous ideas that have held sway for more than a century and to prevent the emergence of new myths".[7] Slattery likewise emphasizes the importance of work done since the opening of those archives.[8]

To help the reader understand this archival data, AG&M introduce their readers to the Holy Office and the Congregation of the Index and to how these bodies functioned. Slattery also gives a very brief introduction to the Congregation. The Holy Office had a broad role regarding matters of faith and morals. The Congregation of the Index, which published the *Index of Prohibited Books* for more than three centuries, was much less important in function and rank. Its decisions were less important; its mission was much more concrete and modest. However, when there was Vatican action in the six cases that AG&M discuss, it was the Congregation of the Index that acted. AG&M emphasize that the previously-mentioned secondary and tertiary writers have often implicated the Holy Office in the proceedings of these cases, yet "the Holy Office played no role in any of these cases".[9]

What is more, while there were indeed "official acts of the Congregation of the Index [not the Holy Office]" in these six cases, there were never "public acts", or at least none recognizable as such. The competence of that Congregation "was ordinarily limited to examining publications and determining whether they should be listed on the *Index of Prohibited Books*". But when the Congregation condemned a book to the *Index*, "the decree of

---

[7] *Negotiating Darwin*, 4-5, 24, 31.
[8] *Faith and Science at Notre Dame*, 7-8, 18-20.
[9] *Negotiating Darwin*, 7-14, 203, 234-235, 270 (quotation), 277, 270. Also *Faith and Science at Notre Dame*, 136-37.

Page 3 of 27

condemnation was all that was made public" and "the reasons for the condemnation were not specified".[10]

The only book to be so condemned for its treatment of evolution was the 1877 book, *New Studies of Philosophy: Lectures to a Young Student* by Fr. Raffaello Caverni, who had served as professor of physics and mathematics in the seminary of Firenzuola. However, Artigas, Glick and Martinez note, "the only decision made public was the prohibition of the book. One might equally conclude that the book was condemned for its critique of the ecclesiastical world, or for the criteria it proposed for scriptural interpretation". Since the book's title made no mention of evolution, "no one not directly involved in the matter would even have suspected an intent to condemn Darwinism". Thus "the condemnation was hardly noticed" and, "when actions of the Holy See adverse to evolutionism have been discussed, the case of Caverni is almost always omitted".[11] Indeed, Brundell and Slattery do not discuss his case. In the other five evolution cases, the Congregation took no public action of any sort.

Brundell and Slattery also note the lack of any clear condemnation of evolution. In Brundell's view, this is despite the best efforts of "reactionary and repressive forces" like the Roman Jesuits, who "appear in a very poor light" in the Vatican-evolution story.[12] Slattery argues that, while AG&M are correct in saying that the Vatican did not issue any public pronouncements on evolution, its private censuring of various authors effectively amounted to a condemnation of evolution, "at least temporarily".[13]

AG&M emphasize from the beginning of their book how in all six of the cases they discuss what is notable is how little action the Vatican took in any of them, and how the Vatican had no set approach, agenda, nor policy regarding evolution. There was no plan at the Congregation of the Index. This emphasis persists to the book's last chapter. Its title is "The Church and Evolution: Was There a Policy?"—a question promptly answered in the chapter's first page: no, there was not.[14]

Both Brundell and AG&M show the overall workings of the Congregation of the Index to be haphazard. In Brundell's view, the Congregation was heavily swayed by Roman Jesuits at the journal *La Civiltà Cattolica*. Yet the Congregation held back from giving full voice to their concerns over evolution out of fear of "giving offence to and rousing opposition from many 'weighty' theologians"—a fear Brundell says was reasonable, given that the Congregation assessed evolution "in a very private process without wide consultation".[15] In other words, the Congregation did not put a solid effort into their evaluation of evolution, and they knew it.

---

[10] *Negotiating Darwin*, 48, 278.
[11] *Ibid.*, 48-50, 51, 279.
[12] "Catholic Church Politics", 83, 92-93.
[13] *Faith and Science at Notre Dame*, 46-47, 160.
[14] *Negotiating Darwin*, 4, 270.
[15] "Catholic Church Politics", 93.

AG&M illustrate the haphazardness of the Congregation of the Index in much greater detail, of course. When there was a complaint against a book, the secretary of the Congregation "was obliged to examine it and to name referees, called 'consultors', to do likewise". Then "a written report was prepared for presentation at a meeting with the consultors and, afterward, at another meeting of the full Congregation of the member cardinals, who composed a definitive resolution submitted for the pope's approval". If a book was found wanting, "a decree was published whereby the book was added to the *Index*".[16]

But new editions of the Index were not issued with any regularity. The consultors and cardinals who were members of the Congregation did not attend meetings regularly: the Congregation consisted of twenty to thirty cardinals during the time covered by *Negotiating Darwin*, but meetings were usually attended by five to ten. There might be multiple reports sought from different consultors, with the views of the consultors not agreeing with each other at all: in the case of Fr. Dalmace Leroy and his 1891 book *The Evolution of Organic Species* (discussed by both AG&M and Brundell), Congregation consultors/referees wrote six different reports over time. The consultors were not in agreement about Leroy's book and not in agreement about evolution. They themselves knew the weakness of the process: one of the consultors writing on Leroy's book suggested that the cardinals not prohibit the book, but rather just warn the author through his superiors to issue a retraction on his own; this would show consideration for the good intentions and the good intellectual and moral qualities of Leroy, "who would not see his book explicitly condemned, while other books like it, that have not been denounced, circulate freely".[17]

It seems that the Congregation of the Index resembled an academic committee, with all the imperfections such a committee entails. Members had other priorities than attending meetings. Reports could be heavily reflective of the views of the referees and thus be inconsistent. The Congregation did not attempt a broad review of books in general, meaning the only books reviewed for listing on the *Index* were the ones about which someone complained. There was no overall plan of action. Given the haphazard nature of the Congregation's workings, it is unsurprising that in five of the six cases discussed in *Negotiating Darwin*, the Congregation took no official action, opting to either privately communicate with authors or to take no action at all.

EVALUATING SCIENCE: AN ANCIENT CASE

Of course, in looking at evolution, the Congregation of the Index was a committee refereeing and communicating about and taking action on (or not taking action on) a matter of science.

---

[16] *Negotiating Darwin*, 9-10.
[17] "Catholic Church Politics", 87; *Negotiating Darwin*, 10-11, 89.

That is why the shadow of Galileo and heliocentrism would have been sensed so heavily by the Vatican as it confronted evolution in the late nineteenth century. But the Galileo case was not the first time the Church had to look at a matter of science and figure out how it related to Scripture. Another case had arisen many centuries before Galileo and many centuries before there was a Congregation of the Index and a Holy Office. This much more ancient case involved science and Genesis 1 and is important for our discussion here.

Genesis 1:14-16 describes the creation of the sun, moon, and stars, and in doing so alludes to their sizes: "And God made two great lights: a greater light to rule the day; and a lesser light to rule the night: and the stars."[18] Were the sky a dome with these lights on its surface, then their "greatness" would be a matter of simple sight. The sun, moon, and stars would all be the same distance from Earth; therefore the relative physical sizes of these bodies would be simply what appears to the eye. The sun and moon would indeed be largest, "greater" than the stars in terms of actual physical bulk.

Careful study of the sky, however, reveals it not to be a dome. Ptolemy (~150 A.D.) in his *Almagest* discussed how the appearance of the stars does not depend on the place on Earth from which they are observed, meaning that the size of the Earth is negligible compared to the distance to the stars:

> Now, that the earth has sensibly the ratio of a point to its distance from the sphere of the so-called fixed stars gets great support from the fact that in all parts of the earth the sizes and angular distances of the stars at the same times appear everywhere equal and alike, for the observations of the same stars in the different latitudes are not found to differ in the least.[19]

The stars of Orion's belt, for example, look no different when observed from Khartoum than from Gdansk. By contrast, observations of the moon from different places on the Earth's globe do differ, showing that Earth is not merely a point in comparison to the distance to the moon (Figure 1).

Ptolemy's science was persuasive. Anyone who travelled and had good eyesight could confirm what he said. Thus, despite the contradiction between the domed universe described by Genesis and the calculations of Ptolemy, St. Severinus Boethius, for example, would cite Ptolemy by name in his *On the Consolation of Philosophy* of 523 A.D. and write of the Earth being but a point compared to the vast stellar distances.[20]

Ptolemy determined the stars to have small but measureable apparent sizes: roughly one fifteenth the apparent diameter of the moon. Observers for fifteen centuries (up until the advent of the telescope) would broadly agree with his measurements. The vast distance to the stars meant that actually they had to be very large in order to appear even as small as they do in the night sky. Ptolemy determined the actual diameter of the most prominent stars to be more than four times that of Earth. He calculated that the sun was actually five times Earth's diameter, while the moon was less than one third of Earth's diameter.[21]

---

[18] Duoay-Rheims translation, appropriate for the time period.
[19] Ptolemy, "The Almagest I, 6" in *Great Books of The Western World (16): Ptolemy, Copernicus, Kepler* (Chicago: W. Benton, 1952), 10.
[20] Philip Ridpath (trans.), *Boethius's Consolation of Philosophy* (London: C. Dilly, 1785), 67.

A prominent star was therefore far "greater" than the moon. Indeed, every visible star in the night sky would greatly exceed the moon in terms of bulk. Anyone with good eyesight who cared to look could at least approximately confirm Ptolemy's measurements of the apparent sizes of the stars compared to the apparent size of the moon. The stars might *appear* small, but the moon *was* small. The moon was arguably not a "great light".

St. Augustine discussed the matter of the "great lights" in his *On the Literal Interpretation of Genesis*, noting that "many of the stars, however, so they [astronomers] boldly assert, are equal to the sun, or even greater, but they seem small because they have been set further away."[22] After further elaboration on what might be said about the celestial lights, Augustine concluded:

> Let them at least grant this to our eyes, after all, that it is obvious that they [sun and moon] shine more brightly than the rest upon the earth, and that it is only the light of the sun that makes the day bright, and that even with so many stars appearing, the night is never as light when there is no moon, as when it is being illuminated by its presence.[23]

St. Thomas Aquinas addressed the "great lights" question in his *Summa Theologica*, Question LXX ("Of the Work of Adornment, as regards the Fourth Day—In Three Articles"). He considers various objections to the Genesis account of the creation of the celestial lights, including,

> *Obj. 5.* Further, as astronomers say, there are many stars larger than the moon. Therefore the sun and the moon alone are not correctly described as the *two great lights*.[24]

His answer to this:

> *Reply Obj. 5.* As Chrysostom says, the two lights are called great, not so much with regard to their dimensions as to their influence and power. For though the stars be of greater bulk than the moon, yet the influence of the moon is more perceptible to the senses in this lower world. Moreover, as far as the senses are concerned, its apparent size is greater.[25]

---

[21] Albert Van Helden, *Measuring the Universe: Cosmic Dimensions from Aristarchus to Halley* (Chicago: University of Chicago Press, 1985), 27, 30, 32, 50

[22] St. Augustine (Bishop of Hippo), *The Literal Meaning of Genesis*, in *The Works of Saint Augustine, a Translation for the 21st Century, Part I, Volume 13: On Genesis: A Refutation of the Manichees; Unfinished Literal Commentary on Genesis; The Literal Meaning of Genesis—introductions, translation and notes by Edmund Hill, O.P., editor John E. Rotele, O.S.A.* (Hyde Park, NY: New City Press, 2002), 211 (Book II, 16.33).

[23] *The Literal Meaning of Genesis*, 212

[24] *"The Summa Theologica" of St. Thomas Aquinas, Part I. QQ. I-LXXIV. Literally Translated by Fathers of the English Dominican Province, Second and Revised Edition* (London: Burns, Oates & Washbourne, Ltd., 1922), 238-239. *"The Summa Theologica"*, 239.

John Calvin made the same general point as Augustine and Aquinas regarding the question of star sizes and Genesis, but at greater length. He praised the findings of astronomers and claimed that Genesis was written in terms of what we see with our eyes, because "The Holy Spirit had no intention to teach astronomy"; "the Spirit of God here opens a common school for all... adapt[ing] his discourse to common usage" and thus Genesis "does not call us up into heaven... [but] only proposes things which lie open before our eyes".[26]

Augustine, Aquinas and Calvin all accepted the science that said that the stars are larger than the moon in terms of actual size, and interpreted Genesis as therefore referring to what our eyes perceive. Others also treated the question of star sizes and Genesis. Some of these, including Robert Bellarmine, were discussing it at the time when heliocentrism was being debated as a scientific idea.[27]

EVALUATING SCIENCE: PARALLELS BETWEEN EVOLUTION AND HELIOCENTRISM

The Church addressed the question of the "great lights" with no Congregation of the Index to weigh in on the matter. The cases of heliocentrism and evolution, however, involved that Congregation's evaluation of scientific questions. There are distinct parallels between those two cases.

There are procedural parallels. When the Vatican confronted heliocentrism in the seventeenth century there was a complaint—about a letter written by Galileo to his friend Fr. Benedetto Castelli concerning a December 1613 conversation that Castelli had regarding heliocentrism with the Grand Duke and Duchess of Tuscany, and with the Grand Duke's heliocentrism-skeptical mother, Christina of Lorraine. The complaint led to consultors and reports.[28] The now-infamous assessment of heliocentrism as "absurd" in a consultors' report of February 1616 was not made public. The only "public act" was by the Congregation of the Index in March 1616. That was not against Galileo but against heliocentrism. Heliocentrism was declared false and contrary to Scripture. Copernicus's *De Revolutionibus* was temporarily prohibited—"suspended until corrected". No specific reasons were given for the condemnation or for why heliocentrism was false.[29]

---

[25] *"The Summa Theologica"*, 242.
[26] The Holy Spirit not teaching astronomy—John Calvin, "Psalm CXXXVI" in *Commentary on the Book of Psalms, by John Calvin, Vol. 5, James Anderson, trans.* (Edinburgh: Calvin Translation Society, 1849) par. 7, 184-185. For the rest, see John Calvin, "Chapter 1" in *Commentaries on the First Book of Moses, called Genesis, by John Calvin, Vol. 1, John King, trans.* (Edinburgh: Calvin Translation Society, 1847) par. 16, 86-87.
[27] Aquinas notes John Chrysostom from the fourth century, although the reference is unclear. Another is Andrew Willet, *Hexapla in Genesin, that is, a Sixfold Commentary upon Genesis, etc.* (London, 1605) who discusses the smallness of the moon versus the stars on page 10. Also see Andreas Cellarius, *Harmonia Macrocosmica, seu Atlas Universalis et Novus* (Amsterdam, 1660), where the "two great lights" question is discussed in the prologue (p. 9, "Solis, & Lunae, duorum Luminarium Magnorum Sacer Textus mentionem facit [etc.]"). Bellarmine touched on the "great lights" question and on Augustine in his Louvain Lectures—see Robert Bellarmine, "Lectures at Louvain (1570-1572): Q. 69. On the Work of the Third Day", in Ugo Baldini and George V. Coyne, S.J., *The Louvain Lectures (Lectiones Lovaniensis) of Bellarmine and the Autograph Copy of his 1616 Declaration to Galileo*; Studi Galileiani Vol. 1, num. 2 (Vatican City: Vatican Observatory Publications, 1984), 22.

But the more interesting parallels between the evolution and heliocentrism cases fall in the area of science. *Negotiating Darwin* emphasizes how the late nineteenth century saw an "eclipse of Darwinism": scientists at that time did not agree on a mechanism for evolution; "Darwin himself, in successive editions of *Origin*... retreated on his claim that natural selection was the sole mechanism of evolution"; "the death knell of Darwinism was sounded at the highest levels of British biology".[30] Darwin's ideas were not so easily confirmed as Ptolemy's stellar sizes and distances. Artigas, Glick and Martinez write that,

> Three outcomes of the eclipse of Darwinism are important to our story. First, the abstract quality of evolutionary discourse in the 1890s made it an obvious target for the kind of Catholic apologetics that we observe in the six cases. Second, the absence of consensus on how competing hypotheses might be tested created a climate propitious for Catholics to advance their own hypotheses.... Third, the lack of agreement among scientists on evolutionary mechanisms gave rise to one of the great shibboleths of religiously based anti-evolutionism, both among Catholics and Protestants: the disagreement of scientists was interpreted as proof of an inherent weakness of the general theory of evolution.[31]

This third item, the supposed weakness of the theory of evolution, appears throughout the works of Brundell, AG&M and Slattery. They reveal one critic after another harping on evolution's perceived *scientific* weakness: Charles Hodge, Presbyterian minister and Princeton professor of theology; Adolphe Tanquerey, the author of widely read theology textbooks; Luigi Tripepi, a consultor for the Congregation of the Index; Enrico Buonpensiere, consultor, rector of the Dominican Studium in Santa Maria Sopra Minerva; G. B. Pianciani, Francesco Salis-Seewis, and Salvatore Brandi, Jesuits writing in the Jesuit publication *La Civiltà Cattolica* (Brandi being "one of the more actively reactionary members of the [*La Civiltà Cattolica*] college of writers" who would come to have great influence over Pope Leo XIII, according to Brundell); the Reverend Jeremiah Murphy, a writer for the *Dublin Review*. Most of these emphasized in some way "the hermeneutic rule that the obvious, natural sense of biblical words should not be abandoned unless it leads to absurd conclusions, which does not happen in this case" because of the scientific problems in the theory of evolution.[32]

---

[28] *The Galileo Affair*, 27-32, 47-48, 134-147.
[29] "Consultants' Report on Copernicanism", in *Ibid.*, 146-147; "Decree of the Index", in *Ibid.*, 149. Note that later events involving Galileo did not include further assessments of heliocentrism.
[30] *Negotiating Darwin*, 20-21.
[31] *Ibid.*, 21.

Salis-Seewis, for example, wrote in *La Civiltà Cattolica* that evolution must first pass its examinations in science's tribunal—"only then will it merit to face Revelation"; "but as long as it presents nothing but the hope of future demonstrations" then it is pointless "to introduce this failure of science in the sacristy, and offer it a chair in hermeneutics".[33] Brandi, a year later, wrote:

> The first impediment to accepting evolution for educated Catholics comes not from the fear of contradicting the Bible, but from the scientific insufficiency of that system, that is, the absolute lack of evidence that confirms it, whether as a theory or as a hypothesis. In this situation, it seems to me that whoever stubbornly defends the theory of the human body's descent from a monkey or any other animal, against the traditional views of the Church Fathers, can with good reason be called rash.[34]

A scientific idea must be solid before it can be used in interpreting Scripture, he said. "It is certainly required", he wrote, "that the words of eternal Truth not be interpreted and warped on the basis of gratuitous hypotheses, to make [those words] say today in obedience to one theory, what will be said tomorrow in obedience to another."[35]

Thus, as regards the rule that the natural sense of the words of Scripture should not be abandoned unless it leads to absurd conclusions: science could show what conclusions were absurd—that had long been clear, thanks to the "two great lights" of Genesis 1; but the science had to be solidly demonstrated. If a theory was absurd scientifically, why bother to consider it theologically? After all, the interpretation of Scripture could not be allowed to simply flutter in the changing winds of passing scientific ideas, following one fallible theory today, another tomorrow.

---

[32] *Faith and Science at Notre Dame*, 58-59 (Hodge); *Negotiating Darwin*, 25 (Tanquerey), 86 (Tripepi), 110 (Buonpensiere), 208 (Salis-Seewis), 225 (Brandi), 245 (Murphy), 150 (quotation); Brundell on Pianciani and on Brandi, "Catholic Church Politics", 83-84, 93-94.

[33] F. Salis-Seewis, "*Evoluzione e Domma* pel Padre J. A. Zahm [etc.]", *La Civiltà Cattolica (1897)* Serie XVI-Vol. X-Quaderno 1118, 201-204: "L'evoluzionismo passi prima i suoi esami al tribunale della scienza, e ne esca colla patente di sistema fondato sopra evidenti principii, deduzioni logiche e fatti positivi, e allora soltanto meriterà di venir messo a confronto colla Rivelazione, se per caso potesse prestare i suoi servigi ad intenderne i sensi, che sarebbe gran gloria della nostra corta ragione umana. Ma finchè egli non presenta altro che la speranza di future dimostrazioni.... finchè le cose stanno così, diciamo , è inutile ed è maraviglia insieme darsi dei cattolici che si fanno premura adesso d'introdurre questo fallito della scienza in sagrestia, e gli offrono una cattedra di ermeneutica". Also see *Negotiating Darwin*, 138-139. Slattery writes in *Faith and Science at Notre Dame* that "one looks in vain for some virtue [148]" in Salis-Seewis's review of Zahm's book, and that "one can safely assume that in his [Salis-Seewis's] mind the entire book was founded on a false sense of empirical evidence and modern science [149]".

[34] *Negotiating Darwin*, 228-229.

[35] *Ibid.*, 228-229.

One perceived scientific weakness of evolution was the matter of infertility of hybrids. A horse and an ass can beget offspring, namely a mule, but that offspring is sterile. This is not considered relevant to evolution today, but Buonpensiere, for example, considered it important.[36] Even Zahm in his 1896 book *Evolution and Dogma* stated that the problem of "the sterility of species when crossed and... the infertility of hybrids" was an argument against evolution that, "as usually advanced appears well-founded".[37]

A concise contemporary presentation of the hybrid infertility argument can be found in Herbert William Conn's 1887 *Evolution of To-day:*

> All the varieties of a species are believed to have descended from the same parent, and can consequently breed with each other. Different species, however, have not thus been connected, and cannot interbreed. Sterility is thus regarded as a bar set by the Creator to prevent the confusion which would result from crossing. It is upon this distinction that those who believe in the stability of species found most of their argument.... Here is a limit beyond which it is claimed variations cannot go. It is pointed out that so far as our observations have gone, no amount of variation has ever produced a new species, if this be admitted as a distinction. In no case has variation produced among domestic varieties any degree of sterility. All of our breeds of pigeons, as widely different as they are, can breed freely with each other; all of the numerous and diverse breeds of dogs are perfectly fertile *inter se*; and so with domestic races in general. Indeed Darwin himself admits that he knows of no well-authenticated case of domestic races being sterile when crossed, a fact which he considers extraordinary, considering the great difference between domestic breeds of dogs, fowls, pigs, etc., and the very slight differences between natural species which are sterile *inter se*. So far, then, as experiments on domestic animals are concerned, the evidence seems to indicate that no amount of ordinary accumulation of variations is able to produce forms so different as to be infertile when crossed. If, then, this be accepted as a criterion for specific distinction, there is no case on record where there has been even an approximation toward the production of a new species.[38]

Another scientific problem, one that *La Civiltà Cattolica* found particularly significant, was the problem of the origin of life. Brundell, whose paper gives much attention to the role of *La Civiltà Cattolica* in the considerations of the Congregation of the Index regarding evolution, highlights this, citing several articles by Salis-Seewis from 1897 with titles including "la generazione spontanea".[39] The first of these articles opens with, "The first postulate with which evolutionism opens its series of imaginative theories is that of the spontaneous generation of the very first organisms." Salis-Seewis then goes on to point out that the idea of the spontaneous generation of life from inanimate matter, while being a very ancient idea, is one that modern science has repudiated. Indeed, he writes, science has repeatedly pronounced judgment on "this first and fundamental supposition" of evolution, and "primordial spontaneous generation has been declared devoid of any foundation and contrary to the constant induction of facts and to one of the best established laws of Nature."[40] Salis-Seewis was correct—while there were still a few advocates for some sort of spontaneous generation even in the late nineteenth century, by the end of the century the idea had been rejected by science, and it remains so today.[41]

---

[36] *Negotiating Darwin*, 93-94. Also "Catholic Church Politics", 88; *Faith and Science at Notre Dame*, 156.
[37] John A. Zahm, *Evolution and Dogma* (Chicago: D. H. McBride & Co., 1896), 182.
[38] Herbert William Conn, *Evolution of To-day: A Summary of the Theory of Evolution as Held by Scientists at the Present Time, and an Account of the Progress Made by the Discussions and Investigations of a Quarter of a Century* (New York: G. P. Putnam's Sons, 1887), 34-35.

The authors of *Negotiating Darwin* state that, given these problems and the eclipse of Darwinism, "evolutionism was viewed by many Catholic theologians as a materialist and agnostic ideology based on a scientific theory that had no serious foundation".[42] Here AG&M see similarity between evolution and heliocentrism/Galileo:

> An evident similarity between the two cases is that both were theories that were not as yet firmly established among scientists. In the case of Galileo, in the seventeenth century, the eleven theologians of the Holy Office whose opinion was sought began by saying [in 1616] that the theory (that the Earth moved) was absurd; once this proposition was introduced, it was easy to dismiss it from the theological point of view. In the case of nineteenth-century evolutionism, theologians presented a whole series of scientific objections, supported by references to, and statements by, scientists, to make the point that it was not logical to base an appeal to theology on such a weak foundation. In both cases, as the scientific arguments grew stronger, theological resistance decreased.[43] That is, what held for evolution and Darwin in the late nineteenth century regarding science and theology held for heliocentrism and Galileo two and one-half centuries earlier. As the changing winds of passing scientific ideas calmed, and as science that once seemed all too obviously fallible began to have the kind of persuasive power possessed by Ptolemy's science of star sizes, scriptural interpretation could be done in light of that science with some confidence of not having today's work of reinterpretation nullified by tomorrow's new theory.

---

[39] "Catholic Church Politics", 88 n. 24. Brundell's highlighting of spontaneous generation is not intentional, as there is no mention of it in the main text of his paper. Also see *Negotiating Darwin*, 70, 131. Spontaneous generation is also touched upon by Slattery, who mentions that Zahm cited it in noting that evolution rests "too much on assumptions and not enough on empirical data" (*Faith and Science at Notre Dame*, 71, 76).

[40] F. Salis-Seewis, "La generazione spontanea e la filosofia antica", *La Civiltà Cattolica* Serie XVI-Vol. XI-Quaderno 1130 (17 July 1897), 142-52, here on 142-43: "Il primo postulato con che l'evoluzionismo apre la serie delle sue imaginose teorie è quello della generazione spontanea dei primissimi organismo…. Qual giudizio abbia ripetutamente pronunziato il tribunal della scienza positiva intorno a questo primo e fondamentale supposto, lo riferiremo appresso più in particolare, chè non è storia da lasciarsi andare in dimenticanza: qui basti il dirne che la generazione spontanea primordiale vi fu dichiarata priva di ogni fondamento e contraria all' induzione costante dei fatti e di una delle meglio accertate leggi della Natura."

[41] Ute Deichmann, "Origin of life. The role of experiments, basic beliefs, and social authorities in the controversies about the spontaneous generation of life and the subsequent debates about synthesizing life in the laboratory," *History and Philosophy of the Life Sciences* 34 (2022), 341–359; John Farley, "The spontaneous generation controversy (1859–1880): British and German reactions to the problem of abiogenesis", *Journal of the History of Biology* 5 (1972), 285-319; S. F. Mason, A History of the Sciences (New York: Macmillan, 1962), 428-431.

[42] *Negotiating Darwin*, 279.

But was anyone at the Vatican really interested in science in the case of heliocentrism? The authors of *Negotiating Darwin* think not. Immediately after making the statement quoted above, they pull back from seeing similarities between the two cases. They state that "there was no calm, objective discussion of heliocentrism" among the Roman authorities in the case of heliocentrism, in contrast to the situation with evolution.[44] This would seem to be an important difference—calm, objective discussion regarding evolution and the case of Fr. Zahm (and others); no such thing for heliocentrism and the case of Galileo.

JESUIT ASTRONOMERS, TYCHO BRAHE, AND HELIOCENTRISM

Of course, Galileo had scientifically proven heliocentrism correct, and opponents of heliocentrism were science deniers who refused to look through a telescope and see that proof for themselves—or so secondary and tertiary writers commonly say.[45] However, these are long-standing erroneous ideas and myths of the same sort that Artigas, Glick and Martinez seek to erase. The ideas of Copernicus were not so easily verified as the claims of Ptolemy. Recent study of the scientific writings of astronomers who opposed heliocentrism has shown that, contrary to what various secondary and tertiary writers have commonly written for well more than a century, the opposition to heliocentrism was objective, calm, and scientifically robust.

Those who opposed heliocentrism focused on a problem involving the sizes of stars in a heliocentric universe—in essence, on a variant of the "great lights" question. Ptolemy discussed how the appearance of the stars does not depend on the place on *Earth* from which they are observed, with the consequence being that the stars are far larger than the *Earth*. In heliocentrism, with the Earth in motion about the sun, the base of observation changes. Now the appearance of the stars does not depend on the place on *Earth's orbit* from which they are observed (the stars of Orion's belt look no different when observed before dawn in October than they do after dusk in March, despite the Earth being in very different places on its orbit in those two months). The consequence now is that the stars are far larger than the *Earth's orbit*; they all utterly dwarf the sun.

---

[43] *Ibid.*, 282.
[44] *Ibid.*, 282.
[45] Christopher M. Graney, "Omission and Invention: The Problematic Nature of Galileo's Proposed Proofs for Earth's Motion," *Logos: A Journal of Catholic Thought and Culture* 22:4 (2019), 78-100; C. M. Graney, "Galileo, a Model of Rational Thinking?" *Catholic Historical Review* 107:3 (2021), 421-429.

This was first pointed out around the turn of the seventeenth century by Tycho Brahe, who also produced a new geocentric model for the universe that, some years later, turned out to be fully compatible with new telescopic discoveries (Figure 2). Some Copernicans, including Johannes Kepler, simply accepted the enormous stars implied by heliocentrism (Figure 3), but Brahe saw the enormous heliocentric star sizes as absurd, and as a strong scientific argument against heliocentrism. Brahe's model generally retained Ptolemy's stellar distances and sizes, and therefore did not suffer from the star size problem.[46]

In the early to mid seventeenth century, following the advent of the telescope, Jesuit astronomers such as Frs. Christoph Scheiner, Giovanni Battista Riccioli, and André Tacquet developed the star size argument further. Scheiner and Tacquet produced brief, elegant versions of this argument (the discussion two paragraphs above, regarding how Earth's orbit in heliocentrism becomes the basis of observation as opposed to the Earth itself, comes from Tacquet).[47] Riccioli set brevity aside. He published large tables containing precise telescopic stellar measurements and the results of calculations made from those measurements, along with pages of discussion—and reached similar conclusions about star sizes.[48]

Thus Brahe's star size argument, and his model, seemed to grow stronger over time, at least until the latter half of the seventeenth century. Robert Hooke in 1674 called the star size argument "a grand objection alledged by divers of the great *Anti-copernicans* with great vehemency and insulting; amongst which we may reckon *Ricciolus* and *Tacquet*… hoping to make it [the Copernican universe] seem so improbable, as to be rejected by all parties."[49] But by that time astronomers including Hooke himself had begun to publish data suggesting problems with measurements of the apparent sizes of stars—problems indicating that such measurements wildly inflated star sizes, even when done carefully and telescopically. Nevertheless, the sizes of stars remained a difficulty for heliocentrism well into the eighteenth century.[50]

The star size argument was known to some of those involved with the Vatican's actions against heliocentrism, both in 1616 and in 1632-33 (following publication of Galileo's *Dialogue*).

---

[46] Dennis Danielson and C. M. Graney, "The Case Against Copernicus", *Scientific American* (January 2014), 72-77; C. M. Graney, *Setting Aside All Authority: Giovanni Battista Riccioli and the Science Against Copernicus in the Age of Galileo* (Notre Dame, IN: University of Notre Dame Press, 2015), 32-37; C. M. Graney, "The Starry Universe of Johannes Kepler," *Journal for the History of Astronomy*, 50:2 (2019), 155-173.

[47] C. M. Graney, *Mathematical Disquisitions: The Booklet of Theses Immortalized by Galileo* (Notre Dame, IN: University of Notre Dame Press, 2017), xviii-xxiii, 30. C. M. Graney, "Galileo Between Jesuits: The Fault is in the Stars," *Catholic Historical Review*, 107 (2021), 197-202.

[48] *Setting Aside*, 129-139.

[49] Robert Hooke, *An Attempt to Prove the Motion of the Earth from Observations* (London, 1674), 26.

[50] *Setting Aside*, 148-157; "Galileo Between Jesuits", 220-224; C. M. Graney, "The Starry Universe of Jacques Cassini: Century-Old Echoes of Kepler," *Journal for the History of Astronomy*, 52:2 (2021), 147-167.

Msgr. Francesco Ingoli, who Galileo believed to have been influential in the rejection of heliocentrism by the Congregation of the Index in 1616, cited the star size argument against Copernicus in his writings. So did Fr. Melchior Inchofer, S.J. who was selected for a three-person Special Commission formed by Pope Urban VIII to investigate the publication of the *Dialogue*.[51]

The star size problem was not the only scientific problem with heliocentrism. There was no physics to explain how the Earth, a sphere of rock and water of obviously vast weight, could be carried around the Sun (Isaac Newton's physics being decades in the future). By contrast, the explanation for how the sun and stars might be carried around the Earth dated to Aristotle—celestial bodies were made of an ethereal substance that moved naturally. Also, heliocentrism called for a rotating Earth. Such rotation should induce deflections in the observed trajectories of projectiles and falling bodies—deflections that were not observed, as Riccioli and other Jesuits (Figure 4) took pains to emphasize.[52] There was, as Brandi would later say about evolution, an absolute lack of scientific evidence to confirm heliocentrism.

A full picture of the role that scientific objections played in the Vatican's actions against heliocentrism is not yet available. More study is needed to better understand the extent to which scientific arguments such as Brahe's, bolstered by the work of astronomers such as Scheiner, motivated those actions and the assessment of heliocentrism as "false". The parallels between the heliocentrism and evolution cases suggest, however, that how the Vatican confronted heliocentrism in the early seventeenth century was similar to how the Vatican confronted evolution in the late nineteenth century—when scientific questions, combined with the idea that the natural sense of biblical words should not be abandoned unless necessary, were significant considerations for the imperfect committee that was the Congregation of the Index.

EVALUATING SCIENCE: WHY?

Why would the Vatican choose to turn loose its imperfect committee-driven processes on the question of heliocentrism? Those investigating the complaint about Galileo's letter to Castelli found no serious fault with the letter. They took no action against Galileo in 1616. However, they chose to delve into the matter of heliocentrism more generally (not that they gave it much time—no more than four days), resulting in the March 1616 "public act" of the Congregation of the Index against heliocentrism (not against Galileo), declaring it false and contrary to Scripture, and temporarily prohibiting Copernicus's *De Revolutionibus*.[53] Why make that choice? Parallels between the evolution and heliocentrism cases may help to answer this question.

---

[51] *Setting Aside*, 66-76; "Galileo Between Jesuits", 214-219.
[52] *Setting Aside*, 29-32, 115-128; Christopher M. Graney and Guy Consolmagno, S.J., "Spin off: The Surprising History of the Coriolis Effect and the Jesuits Who Investigated it", *Catholic Historical Review* 109:2 (2023), 302-320. The Coriolis Effect is a deflection in the observed trajectories of projectiles and falling bodies induced by Earth's rotation. Jesuit scientists in the seventeenth century correctly envisioned it occurring on a rotating Earth, but the Effect, although it does exist, is surprisingly difficult to detect, and they failed to detect it.

The reason given by the Congregation of the Index in 1616, and re-iterated in the sentence pronounced against Galileo in 1633, was to prevent the "false" doctrine of heliocentrism from advancing further against the truth—a doctrine "altogether contrary to Holy Scripture"; a "pernicious" doctrine containing "various propositions against the authority and true meaning of Holy Scripture". The action of 1616 was taken to "completely eliminate" heliocentrism, and to "remedy the disorder and the harm which derived from it and which was growing to the detriment of the Holy Faith".[54]

Yet the long-standing matter of Genesis and the "two great lights" was a template for addressing heliocentrism. Thanks to the science of Ptolemy, the Bible had long been taken as speaking to appearances regarding the apparent *sizes* of celestial bodies. That logic could certainly be applied to the science of Copernicus and the apparent *motions* of celestial bodies. It was not. Again, why that choice?

In Brundell's and AG&M's discussions of evolution and the Vatican, here and there they touch upon something that suggests an answer. In their work we see various people emphasizing the importance of the descent of all people from Adam, and the unity of humankind: a provincial council; a bishop; Jesuit critics; a Pontifical Biblical Commission.[55]

Bishop John Cuthbert Hedley of Newport, Wales, for example, in an 1898 review of four works by Zahm, wrote that "there are some matters so clearly revealed as to be out of the field of question or investigation. There is, for example, the point of the unity of the human race, as Dr. Zahm himself admits." This unity, Hedley thought, was one of several subjects, "in which it would be not only a mistake, but also an offence against religious faith, not to start with a firm hold of what is taught by the Church—taught, that is to say, indirectly, and implied in theological dogma."[56] As another example, Brundell notes that "The Pontifical Biblical Commission in 1909 expressed no repugnance for evolution as such but was concerned about the historicity of the Genesis account of the origins of the human race.... The Commission could not see how some aspects of the first chapters of Genesis could not be literally historical, including... the necessity for a monogenistic origin of the human race as the basis for its unity."[57]

Genesis, read plainly, was clearly "monogenistic". That is, it supported the idea of a single origin for all humanity. All human beings are descendants of the same parents, Adam and Eve, and thus of one family.

---

[53] *The Galileo Affair*, 29-30, 146-50; Annibale Fantoli, *Galileo: For Copernicanism and for the Church*, 3rd. ed.; Studi Galileiani Vol. 6 (Vatican City State: Vatican Observatory Publications, 2003), 168-78.
[54] *The Galileo Affair*, 149, 288-89.
[55] e.g. *Negotiating Darwin*, 23, 65, 226; "Catholic Church Politics", 86 with n. 14 & 16 and 94 with n. 50.
[56] *Negotiating Darwin*, 226.
[57] "Catholic Church Politics", 94 with n. 50.

A competing idea, however, was that of separate origins for different "types" of people—for example, light-skinned, light-haired Europeans and dark-skinned, dark-haired Africans. There were, according to this idea, actually different species of human (-like) creatures, with these different species commonly being called "races". This is "polygenism"—multiple origins for the multiple human "types" or "races", with most "races" not being of the line of Adam and Eve.

Polygenism was an ancient idea. It had been bolstered in European minds by voyages of discovery that revealed distant, peopled lands. Polygenism was also considered to be very much heretical.[58]

A remarkable intersection between polygenism and heliocentrism can be found in Giordano Bruno, a Copernican who argued for an infinite universe of other suns orbited by other inhabited Earths. Bruno argued in 1591 that the different "races" could not all have a common origin:

> For of many colors
> Are the species of men, and the black race
> Of the Ethiopians, and the yellow offspring of America...
> Cannot be traced to the same descent, nor are they sprung
> From the generative force of a single progenitor.

Bruno noted that "it is said in the prophets... that all races of men are to be traced to one first father", but he adds that "no one of sound judgement can refer the Ethiopian race to that protoplast."[59] David Livingstone has argued that Bruno's thinking here was "of a piece with his belief... in the existence of other planets with inhabitants evidently not traceable to Adam".[60]

Bruno's singling out "Ethiopians" was typical. For those Europeans who believed that there were different races/species of people, the race/species that was typically considered most removed from Adam's lineage was the "black race", as "sound judgement" supposedly indicated.

Of course it was countered that sound judgement, and indeed science, would say otherwise. Morgan Godwyn pointed out in his 1680 book *The Negro's & Indians Advocate, Suing for their Admission into the Church: or Persuasive to the Instructing and Baptizing of the Negro's and Indians in our Plantations* that different species do not beget fertile offspring. Noting the sterility of mules (discussed earlier in this paper), Godwyn wrote that, if different races were different species, then the "*Mulatto's* and *Mestizo's*.... and all these *Moors*, must, *like the Mules*... be for ever Barren". They would all lack the "prolific faculty, and never be able to procreate their like". And yet "the contrary whereof is daily seen"—all these "mixed race" people are fertile; they have children.[61] Thus, it would seem that humans are of one family, whatever "race" they may be, and that was just a fact of science.

---

[58] David N. Livingstone, *Adam's Ancestors: Race, Religion & the Politics of Human Origins* (Baltimore: Johns Hopkins University Press, 2008), 1-51, especially 51 regarding being "heretical".
[59] J. S. Slotkin (ed.), *Readings in Early Anthropology* (Chicago: Aldine Publishing Co., 1965), 43.
[60] *Adam's Ancestors*, 23-24.

But the polygenists claimed science anyway. For example, J. H. Van Evrie (M.D.) in his 1861 book *Negroes and Negro "Slavery"* argued that not all "hybrids" had the immediate infertility of a mule:

> [T]he inference... that whites and negroes were of the same species, because the mulatto, unlike the mule, did reproduce itself, is simply absurd.... The mulatto, literally speaking, or in the ordinary sense, does beget offspring, but mulattoism is as positively sterile as muleism. The phenomenon of hybridity is manifested, as has been stated, in conformity with the nature of the beings concerned, and as the human creatures are separated by an almost measureless as well as impassable distance from the horse and ass, the laws of hybridity are, of course, correspondingly different. Instead of a single generation, as in the animals referred to, sterility in the human creatures is embraced within four generations, where a boundary is arrived at as absolutely fixed and impassable as the single generation in the case of the former.[62]

Van Evrie went on to argue that the evidence for this eventual sterility was so overwhelming that all slave traders, and many of those who held people in slavery, knew of it from casual observation, without being aware that anyone had "formalized the essential character of mulattoism".[63]

Van Evrie was not alone in his ideas. The 1850 meeting of the American Association for the Advancement of Science featured discussion of how the "types" of human beings were fixed, because "hybrids" were sterile in the long term, and thus died out.[64] Varieties within a species, being descended from the same parents, can consequently breed with each other. Different species, being not so descended, cannot.

The science of polygenism was central to the whole business of racial slavery and oppression. As Van Evrie himself noted,

> If the Negro had descended from the same parentage, or, except in color merely, was the same being as ourselves.... then it would be [a Christian's] first and most

---

[61] Morgan Godwyn, *The Negro's & Indians Advocate, Suing for their Admission into the Church: or Persuasive to the Instructing and Baptizing of the Negro's and Indians in our Plantations* (London, 1680), 23-24.
[62] J. H. Van Evrie, *Negroes and Negro "Slavery": The First an Inferior Race; the Latter Its Normal Condition* (New York: Van Evrie, Horton & Co., 1863), 145-146.
[63] *Negroes and Negro "Slavery"*, 147.
[64] Guy Consolmagno, S.J. and Christopher M. Graney, *When Science Goes Wrong: The Desire and Search for Truth* (New York: Paulist Press, 2023), 102-103; see "An examination of the Physical History of the Jews, in its bearings on the Question of the Unity of the Races; by JOSIAH C. NOTT, M.D., Mobile, Ala." and the affirmative response from Prof. Louis Agassiz, in *Proceedings of the American Association for the Advancement of Science, Third Meeting, Held at Charleston, S.C., March 1850* (Charleston: Walker and James, 1850), 98-107.

> imperative duty... to set an example to others, to labor night and day to elevate this (in that case) wronged and outraged race—indeed, to suffer every personal inconvenience, even martyrdom itself in the performance of a duty so obvious and necessary.[65]

This sentiment was not limited to Americans tied up with the question of slavery. The autonomy of science was cited, and the shadow of Galileo invoked, by many others who argued that traditional religious ideas about the unity of and nature of human beings should yield to scientific evidence that showed that there were different species of human (-like) beings.[66] An issue of *The Anthropological Review* from the mid-nineteenth century provides illustration of all this. In this publication of the Royal Anthropological Institute of Great Britain and Ireland we find, by Georges Pouchet of the Muséum national d'Histoire naturelle in Paris:

> The battle fought and won by astronomy in the days of Galileo, was in truth but the beginning of the war, and alone would have proved utterly inadequate to teach men of science their strength, or theologians their weakness. This was shown in the reception accorded to geology, whose stupendous revelations from the page of nature were long expected to bend to a written record. It is still shown in the criticisms provoked in certain quarters by anthropological investigations. We are free to speculate on the age of rocks, and even to inquire into the succession of plants and animals; but man is a sacred, and, therefore, a forbidden subject. His origin, antiquity, and special relationships have all been settled by a tribunal that laughs at induction, and treats opposing facts with derision.... On special difference as attaching to brown and white bears, and of organic diversity in relation to African and Asiatic elephants, it was quite lawful to dilate, but an Esquimaux and a European, a Negro and a Persian, were to be invariably treated as of one species....[67]

Pouchet proceeds to discuss, "the inferior races, more especially the Negroid" and, "the true man, by which we mean the large-brained and small-mouthed Caucasian".[68] Another article in the issue goes on about "the Negroid type" being incapable of building a civilization such as Egypt, even had they tens of thousands of years to do so. It notes that,

> The time has now assuredly come, when the accepted fallacies of a learned barbarism should succumb to the clear demonstrations of inductive science, and racial facts be championed to their appropriate place, as among the most

---

[65] *Negroes and Negro "Slavery"*, 59-60.
[66] David N. Livingstone, *Adam's Ancestors: Race, Religion, and the Politics of Human Origins* (Baltimore: Johns Hopkins University Press, 2008); Ibram X. Kendi, *Stamped from the Beginning: The Definitive History of Racist Ideas in America* (New York: Nation Books, 2016), 50-51, 133-138, 179-209.
[67] Georges Pouchet, "The Plurality of the Human Race, translated and edited by H. J. C. Beavin," *The Anthropological Review* 3:9 (May 1865), 120-121; Consolmagno and Graney, *When Science Goes Wrong*, 102-108.
[68] "The Plurality of the Human Race", 126-127.

important and reliable data upon which history, more especially that of the earlier ages, can be based.[69]

Of course, ideas such as these are not part of science today. Modern science is monogenistic. It says that all people are of the same family, and that "racial" variations are quite minor, compared to the variations between individuals, and compared to variations found within other species.

Indeed, the sorts of ideas promoted by Van Evrie and Pouchet have been so thoroughly rejected that today they are termed "pseudo" science. Today, no scientist uses the language of their "scientific racism". By contrast, the language of geocentrism, of "sunrise" and "sunset", remains common, even among astronomers. To paraphrase Hedley, today it is considered not only mistaken, but offensive, not to start with a firm hold of the idea that all human beings, regardless of their "race", are of the same family, and fundamentally equal. Were a scientist today to propose a new polygenistic theory for the origin of human beings, asserting that certain types of people are not truly of the human family, that theory, and that scientist, would be roundly condemned.[70] As Hedley said, some matters are considered to be so clearly revealed as to be out of the field of question or investigation. Today we simply reject the idea that science can tell us that the man or woman standing next to us is not fully human. Those who do not reject it are ranked among the least pleasant kinds of crackpots.

Therefore, insofar as evolution and polygenism and the idea that certain people were not fully human were all linked together in the nineteenth century, even people today who care little for the Catholic Church might understand why the Vatican would choose to turn loose its imperfect committee-driven processes on the evolution question. Even people today who have little interest in discussions of original sin and salvation history will understand why the unity of humankind must be sacrosanct. The example of "scientific racism" urges that science be subject to confrontation and criticism from outside of science.

Given this, what is the parallel in the case of heliocentrism? What was sacrosanct in that case? It seems that what was sacrosanct was reinterpreting Scripture only when *necessary*.

We have seen from the "two great lights" case that Augustine, Aquinas and Calvin all accepted the need to reinterpret Scripture in the light of scientific evidence. Of these three, only Calvin (1509-1564) lived to see the advent of heliocentrism. Further exploration of his writings is instructive here.

In the case of the two great lights, Calvin gave a spirited defense of astronomy, even as it contradicted a plain reading of Genesis 1:14-16:

> Moses wrote in a popular style things which without instruction, all ordinary persons, endued with common sense, are able to understand; but astronomers investigate with great labor whatever the sagacity of the human mind can comprehend. Nevertheless, this study is not to be reprobated, nor this science to be condemned, because some frantic persons are wont boldly to reject whatever is unknown to them. For astronomy is not only pleasant, but also very useful to be known: it cannot be denied that this art unfolds the admirable wisdom of God. Wherefore, as ingenious men are to be honored who have expended useful labor on this subject, so they who have leisure and capacity ought not to neglect this kind of exercise.[71]

---

[69] "Race in History," *The Anthropological Review* 3:9 (May 1865), 238-239.
[70] *When Science Goes Wrong*, 108.

Yet despite his admiration for astronomy, Calvin rejected heliocentrism:

> We will see some who are so deranged, not only in religion but who in all things reveal their monstrous nature, that they will say that the sun does not move, and that it is the earth which shifts and turns. When we see such minds we must indeed confess that the devil posses them, and that God sets them before us as mirrors, in order to keep us in his fear. So it is with all who argue out of pure malice, and who happily make a show of their imprudence. When they are told: "That is hot," they will reply: "No, it is plainly cold."[72]

Calvin's logic regarding the two great lights could certainly be applied to heliocentrism, so apparently he simply was unpersuaded by heliocentrism. To him, it was merely some baseless hypothesis. To reinterpret Scripture to accommodate it would be religiously deranged, or devilish.

Echoes of this can be found in the seventeenth century among those who interacted with Galileo. When Galileo queried Cardinal Carlo Conti in 1612, Conti offered the opinion that an orbiting Earth was not consistent with Scripture, and that therefore heliocentrism could only be reconciled with Scripture by invoking the idea that the Bible was speaking according to common usage of language. But, Conti warned, "*that* mode of interpretation is not to be admitted unless absolutely necessary".[73]

Likewise Cardinal Robert Bellarmine wrote a few years after Conti:

> If there were a true demonstration that the sun is at the center of the world and... does not circle the earth but the earth circles the sun, then one would have to proceed with great care in explaining the Scriptures that appear contrary, and say rather that we do not understand them than that what is demonstrated is false. But I will not believe that there is such a demonstration,

---

[71] "Chapter 1" in *Commentaries on the First Book of Moses,* 86-87.
[72] John Calvin, "Huitieme Sermon: 1. Cor. Chap. X, v. 19-24", in *Ioannis Calvini opera quae supersunt omnia*, XLIX, G. Baum, E. Cunitz, E. Reuss, eds. (Brunsvigae: apud C. A. Schwetschke et Filium/Appelhans & Pfenningstorff, 1892), 671-684, on 677; translation by Robert White in "Calvin and Copernicus: the Problem Reconsidered", *Calvin Theological Journal* 15 (1980), 233-243, at 236-237. See also John Hedley Brooke, *Science and Religion: Some Historical Perspectives* (Cambridge: Cambridge University Press, 1991), 96-97.
[73] Galileo Galilei, Christoph Scheiner, *On Sunspots, translated and introduced by Eileen Reeves and Albert Van Helden* (Chicago: University of Chicago Press, 2010), 351. See also *Galileo: For Copernicanism*, 116.

> until it is shown me.... I have very great doubts about [available demonstrations], and in case of doubt one must not abandon the Holy Scripture as interpreted by the Holy Fathers....[74]

Bellarmine had applied this logic to his own ideas. He had once argued against the prevailing view of astronomers that said that the stars were carried around the Earth together on a solid celestial sphere. Various scriptural verses, Bellarmine said, suggested that the stars rather moved autonomously yet in lock step, like a school of fish in the sea or a formation of birds in the air. But, he said, should science eventually show that stars were carried on a sphere, the interpretation of Scripture would have to be adapted to the science:

> If then one ascertained with evidence that the motions of the heavenly bodies are not autonomous… one would have to consider a way of interpreting the Scriptures which would put them in agreement with the ascertained truth: for it is certain that the true meaning of Scripture cannot be in contrast with any other truth.[75]

A third example in addition to Conti and Bellarmine is Riccioli. He argued in his 1651 *New Almagest* that,

> If the liberty taken by the Copernicans to interpret scriptural texts and to elude ecclesiastic decrees is tolerated, then one would have to fear that it would not be limited to astronomy and natural philosophy and that it could extend to the most holy dogmas; thus, *except in cases of manifest necessity*, it is important to maintain the rule of interpreting all sacred texts in their literal sense.[76]

Riccioli then proceeds to state that science shows that there is no such necessity. Indeed, much of the *New Almagest* was dedicated to showing that there was no manifest necessity, because based on "reasoning and intrinsic arguments alone", and with "every authority set aside", heliocentrism was false and inconsistent with what was known from physics, astronomy, and mathematics[77]—as seen from, for example, Riccioli's aforementioned tables of stellar measurements, calculations, etc.

---

[74] "Cardinal Bellarmine to Foscarini (12 April 1615)", in *The Galileo Affair*, 68.

[75] Robert Bellarmine, "Lectures at Louvain", 20. Also see Richard J. Blackwell, *Galileo, Bellarmine, and the Bible* (Notre Dame, Indiana: University of Notre Dame Press, 1991), 39-43. Bellarmine is willing to adapt the interpretation of Scripture to what is ascertained through science, but his grasp of the science is not always strong. When he touches on the "two great lights" question in his Louvain Lectures ("Lectures at Louvain", 22) he notes that "astrologers" say the stars are much more distant than the moon, but he expresses doubt whether that is true. He apparently does not grasp Ptolemy's argument and the straightforward observations and calculations that underpin it. His doubt about demonstrations in science expressed in his 1615 letter probably stems from his not following them, and thus being unpersuaded by them.

[76] Italics added. Some writers, in translating Riccioli's words, have omitted the italicized words, so that Riccioli seems to be arguing that the literal meaning of biblical statements always takes precedence over science. See *Setting Aside*, 244-245.

[77] *Ibid.*, 245, 5-6, 162.

Riccioli does not specify what most holy dogmas he has in mind, but of course the dogma of the unity of humankind comes to mind at this point. "Are we to tolerate the followers of Bruno regarding heliocentrism?" we can imagine Riccioli saying; "If so, what will we do when they start pushing Bruno's ideas about Ethiopians not being true people?"

CONCLUSION

Today we do not accept what science in the past had to say about who were and were not true people, fully human. Today we would not accept claims supposedly based in science about who are and are not true people. Thus, there exists a subject that is considered to be, in essence, sacred, and out of the field of investigation. The past science that once purported to investigate that subject and made such claims we now call "pseudo-science".

We see in the Vatican's confrontation of evolution in the late nineteenth century an effort to wrestle, using a very imperfect method, with science that was unsettled and consequential. Evolution was consequential—at least insofar as it was seen then as allowing for a polygenistic origin of humankind (or, as might have been seen at the time, human*kinds*), thus undermining the unity of humanity and allowing for some people to be considered less than fully human. Evolution was unsettled—Darwinism was in "eclipse", and many thinkers saw serious scientific questions associated with it. At least one of these, the question of the origin of life, remains unanswered today. The Vatican's process for wrestling with the idea of evolution was, in essence, a committee of men who lacked the time, expertise and commitment really necessary to address the matter at hand.

The evolution case suggests a similar situation regarding the Vatican's confrontation of heliocentrism in the early seventeenth century. Heliocentrism was unsettled at the time—the question of the absurdly large stellar sizes that heliocentrism implied at that time was widely recognized as a powerful anti-Copernican scientific argument. Heliocentrism certainly seems less consequential to modern eyes than evolution and the sorts of unity-of-humanity questions associated with evolution in the nineteenth century. It seems that the question that was considered to be consequential at the time was not whether scriptural interpretation could be accommodated to heliocentrism, but whether it should be so accommodated absent manifest necessity. Scriptural interpretation had long been accommodated to science when necessary, as seen in the case of Genesis 1 and the sizes of the "two great lights". But to let scriptural interpretation flutter in the changing winds of unsettled and obviously fallible science would put at risk far more consequential ideas than Earth's fixity—ideas like, perhaps, the idea of the unity of humankind. And of course, these matters were addressed by a very imperfect committee.

The Vatican's processes in the heliocentrism and evolution cases were imperfect, but it is difficult to envision better processes, or to envision no processes. These cases may have become emblematic of "the problems between science and religion", but it is easy to imagine them becoming emblematic of the problems between science and *not* religion. If today a new scientific idea arose that was promising yet very much unverified, and that had implications for who was fully human and who was not, with divisions along lines of "race", things similar to the Vatican's actions would likely take place; the Vatican would likely not be the dominant player in them. Panels and committees would be formed; reports would be issued; harsh words would be said. There would be consequences to individuals at least as great as what happened to Fr. Zahm (who was privately asked to retract his book *Evolution and Dogma*). The process would be imperfect. Modern processes for dealing with consequential scientific ideas, whether they involve the development of weapons or the response to deadly diseases, have been imperfect.

Hopefully they would not be as imperfect as the process brought to bear against Galileo in 1633. He was sentenced to prison and then perpetual house arrest. He had the misfortune to run afoul of a powerful man, Pope Urban VIII, who once had the birds in the papal garden killed when their noise came to bother him.[78] Urban had once addressed Galileo "as a brother" and had written poetry praising his telescopic discoveries; before the publication of the *Dialogue*, Urban's powerful nephew had said that Galileo had "no better friend" than Urban, and Urban had granted Galileo an audience. After publication, however, Urban came to explode into anger at Galileo's name.[79] But modern leaders are also imperfect, as are modern processes that might protect modern Galileos from abuses of power.

Despite the abuse brought to bear against Galileo, the Vatican failed to "completely eliminate" heliocentrism. Indeed, heliocentrism prevailed completely. The Earth circles the sun. Scripture has been reinterpreted to accommodate.

But the situation is different with evolution. In many ways, evolution has prevailed much like heliocentrism did. A striking example of the Catholic Church reinterpreting Scripture to accommodate an evolutionary view of the universe is the proclamation of "The Nativity of Jesus Christ from the Roman Martyrology", often recited during the celebration of the Liturgy of the Hours on December 24 and before Midnight Mass at Christmas. Traditionally, this text stated that Christ was born in "the year from the creation of the world, when in the beginning God created heaven and earth, five thousand one hundred and ninety-nine".[80] Today, the text states that Christ was born "when ages beyond number had run their course from the creation of the world".[81] But the evolution that has prevailed is a monogenistic evolution. The unity of man is preserved. That "most holy dogma" also prevailed.

---

[78] James Lees-Milne, *Saint Peter's: The Story of Saint Peter's Basilica in Rome* (Boston: Little, Brown & Co., 1967), 248; Mariano Artigas and William R. Shea, *Galileo in Rome: The Rise and Fall of a Troublesome Genius* (New York: Oxford University Press, 2003), 136.

[79] *Galileo in Rome*, 139-141, 172-176; *The Galileo Affair*, 38, 229-230 (5 September 1632 letter of Francesco Niccolini, Tuscan ambassador to Rome).

[80] Prosper Guéranger, *The Liturgical Year: Advent. 2d ed.* (Dublin: James Duffy, 1870), 536; Cesare Baronio, *Martyrologivm Romanvm Ad nouam Kalendarij rationem, & Ecclesiasticae* (Venice, 1602), 695-696.

It prevailed not through some Vatican decree intended to protect it and to "completely eliminate" polygenism and scientific racism. We might wish that some decree could have squelched those ideas and remedied "the disorder and the harm" that derived from them. They and their offspring, eugenics, thrived for decades, to the detriment of many, especially those with the least power.[82] Those with the least power needed the Vatican. They needed the Congregation of the Index.

The history of the Vatican's efforts to confront evolution reflects the need for a process, a committee, a Congregation, even if imperfect, for confronting fallible science. The history of the Vatican's efforts to confront heliocentrism reflects the need for vigilance in ensuring that process is not abused. Both histories need an understanding of the Church's much older confrontation with the matter of the "two great lights" of Genesis 1.

---

[81] United States Conference of Catholic Bishops, "The Nativity of our Lord Jesus Christ from the Roman Martyrology", https://www.usccb.org/prayer-and-worship/liturgical-year-and-calendar/christmas/christmas-proclamation (accessed 22 Feb. 2024).
[82] *When Science Goes Wrong*, 100-106.

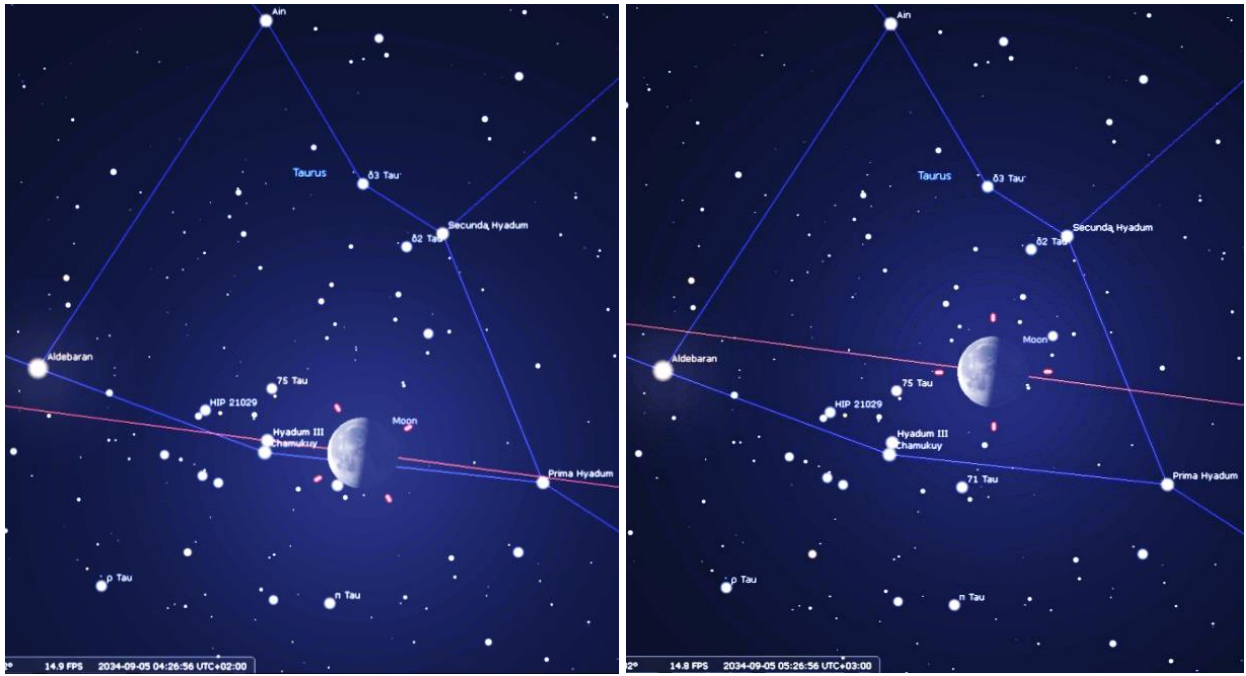

**Figure 1: The moon passing through the Hyades in Taurus on 5 September 2034, as seen from Gdansk (left) and Khartoum (right). The red line is the moon's path through the stars. Observations of the same stars from different latitudes do not differ, indicating that the stars are so distant that the size of the Earth is as nothing by comparison, but this is not true in the case of the moon. Images made with the planetarium software *Stellarium*.**

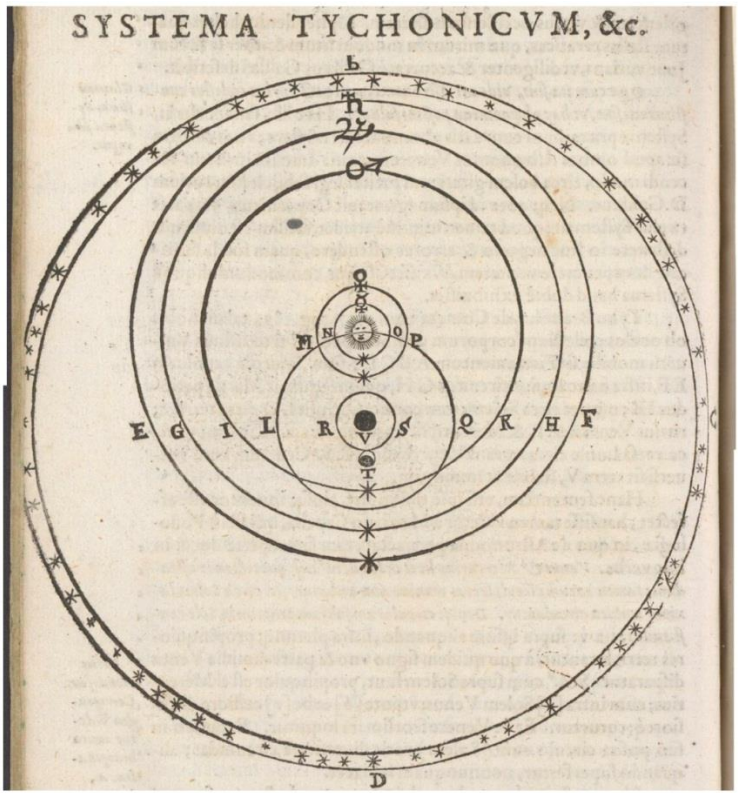

**Figure 2: A diagram from the 1614 *Mathematical Disquisitions* of Christoph Scheiner, S.J. and his student Johann Georg Locher, showing the geocentric model of the universe proposed by the astronomer Tycho Brahe. The planets circled the sun; the sun, moon, and stars circled the immobile Earth. This model was indistinguishable from the Copernican model in terms of astronomical observations made from Earth. Thus it was as compatible with new telescopic discoveries as was the Copernican model.**

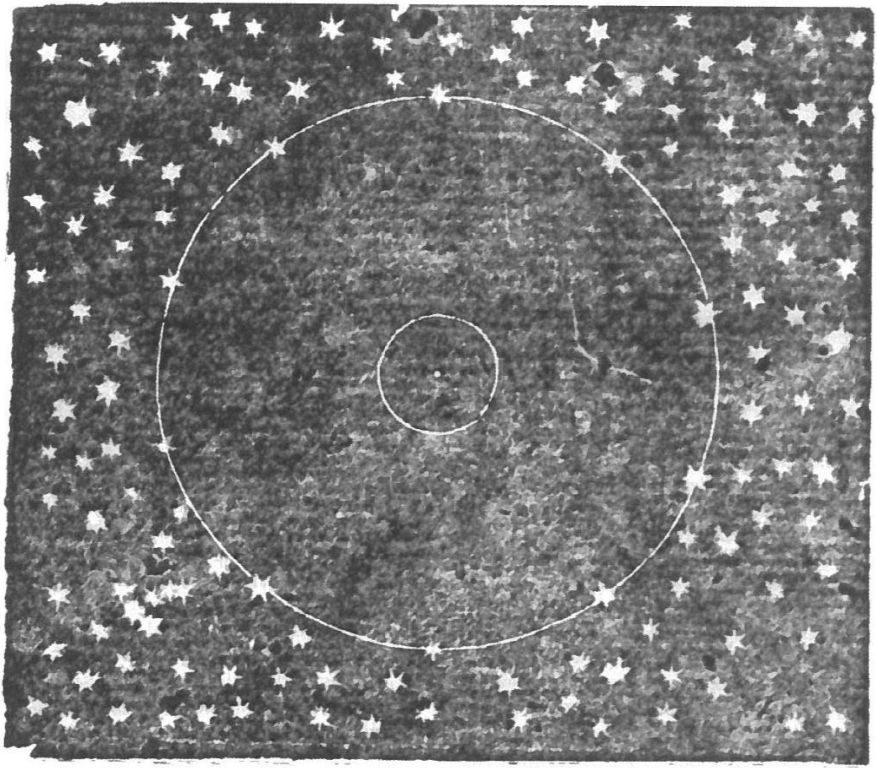

Figure 3: Diagram of the universe of stars from Johannes Kepler's 1618 *Epitome of Copernican Astronomy*, showing a small sun (the dot at the center) surrounded by large stars.

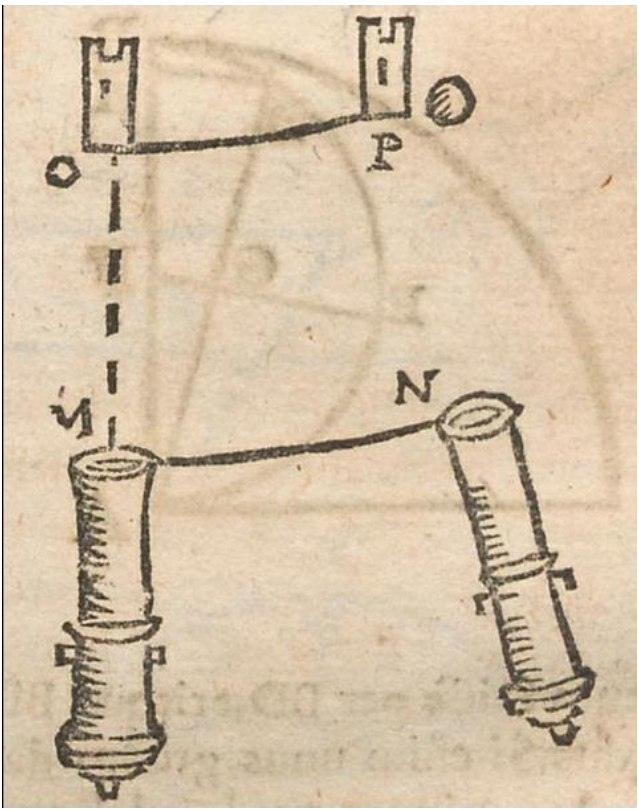

Figure 4: Diagram from a section on "Objections against Copernicus" in the 1674/1690 book *Mathematical World* by Claude François Milliet Dechales, S.J. The diagram shows a cannon fired toward a target on a rotating Earth. The Earth's rotation, in carrying the cannon and the target to the right by differing amounts on account of their differing latitudes and the spherical shape of the Earth, causes the cannon ball's trajectory to deflect, such that it passes wide of the mark at P. This deflection effect does exist and is today known as the "Coriolis Effect", but it had not been observed in the seventeenth century and its absence was advanced by various Jesuit astronomers as an argument against heliocentrism.